\documentclass[prb,aps,twocolumn,groupedaddress,floats,final,nopacks]{revtex4-2}
\usepackage{graphicx}
\usepackage{subfigure}
\usepackage{amssymb}
\usepackage{latexsym}
\usepackage{epsfig}
\usepackage{psfrag}
\usepackage{dcolumn}
\usepackage{amsmath,amsfonts,amssymb,bm,ulem}
\usepackage{color}
\usepackage{float}

\newcommand{\be}{\begin{equation}}
\newcommand{\ee}{\end{equation}}
\newcommand{\bea}{\begin{eqnarray}}
\newcommand{\eea}{\end{eqnarray}}

\begin{document}
\title{Landau Damping in an Electron Gas}

\author{I.S. Tupitsyn}
\affiliation{Department of Physics, University of Massachusetts, Amherst, MA 01003, USA}
\author{N.V. Prokof'ev}
\affiliation{Department of Physics, University of Massachusetts, Amherst, MA 01003, USA}


\begin{abstract}
Material science methods aim at developing efficient computational schemes for describing complex many-body effects and how they are revealed in experimentally measurable properties. Bethe-Salpeter equation in the self-consistent Hartree-Fock basis is often used for this purpose, and in this paper we employ the real-frequency diagrammatic Monte Carlo framework for solving the ladder-type Bethe-Salpeter equation for the 3-point vertex function (and, ultimately, for the system's polarization) to study the effect of electron-hole Coulomb scattering on Landau damping in the homogeneous electron gas. We establish how this damping mechanism depends on the Coulomb parameter $r_s$ and changes with temperature between the correlated liquid and thermal gas regimes. In a broader context of dielectric response in metals, we also present the full polarization and the typical dependence of the exchange-correlation kernel on frequency at finite momentum and temperature within the same computational framework.
\end{abstract}

\maketitle

\section{Introduction}

Landau damping (LD) was originally introduced to describe decay of charge oscillations in plasma \cite{Landau1946}. Its importance is hard to overestimate because it appears in multiple physics contexts from electron liquids in metals to nuclear matter, galactic dynamics \cite{LB1962}, and interstellar plasma turbulence \cite{Howes2019} (for reviews see Refs.~\cite{Ryutov1999,Mendonca2023}). In condensed matter physics, LD mechanism is most often associated with energy losses due to creation of electron-hole pair excitations \cite{mahan3rd} at frequencies $\Omega$ below $v_F Q $ where $Q$ is the pair momentum and $v_F$ is the Fermi velocity.

The standard way of characterizing energy losses in the homogeneous electron gas (HEG) due to interactions between the charges is through the imaginary part of the inverse dielectric function $\epsilon^{-1}(Q,\Omega )$, or, equivalently, the system's polarization $\Pi(Q,\Omega )$; the two quantities are straightforwardly related by $\epsilon = 1-V\Pi $, where $V(Q)=4\pi e^2/Q^2$ is the Coulomb potential. In what follows we define the LD coefficient as the quantity controlling the slope of the linear low-frequency dependence of $\textrm{Im}\Pi$
\begin{equation}
\textrm{Im}\Pi (Q,\Omega) = - \gamma_{\textrm{LD}} \, \frac{\Omega}{v_F^{(0)} Q} ,
\quad \Omega \ll v_F^{(0)} Q \ll \varepsilon_F .
\label{LDdef}
\end{equation}
Here $v_F^{(0)}=k_F/m$ and $\varepsilon_F = k_F^2/2m$ are the Fermi velocity and Fermi energy of the ideal gas, respectively, and $m$ is the electron mass. In the non-interacting system at zero temperature the LD coefficient is proportional to the Fermi surface density of states, $\gamma_{\textrm{LD}} = (\pi /2) \rho_F^{(0)}$, with $\rho_F^{(0)} = mk_F/\pi^2$. For convenience, below we measure momenta in units of $k_F$ and energies in units of $\varepsilon_F$. By Luttinger theorem, the electron number density is given by $n=k_F^3/3\pi^2$.

The HEG model describes electrons interacting via the long-range Coulomb force on a positively charged neutralizing background. The corresponding Hamiltonian is defined by
\begin{equation}
H=\sum_i \frac{k_i^2}{2m} + \sum_{i<j} \frac{e^2}{|\mathbf{r}_i -\mathbf{r}_j|} - \mu N .
\label{jellium}
\end{equation}
The strength of many-body correlations is characterized by the Coulomb parameter $r_s=(4\pi a_B^3 n /3)^{-1/3}$,
which measures inter-particle distance in terms of the Bohr radius $a_B=1/me^2$.


\begin{figure}[H]
\centerline{\includegraphics[angle = 0,width=0.99\columnwidth]{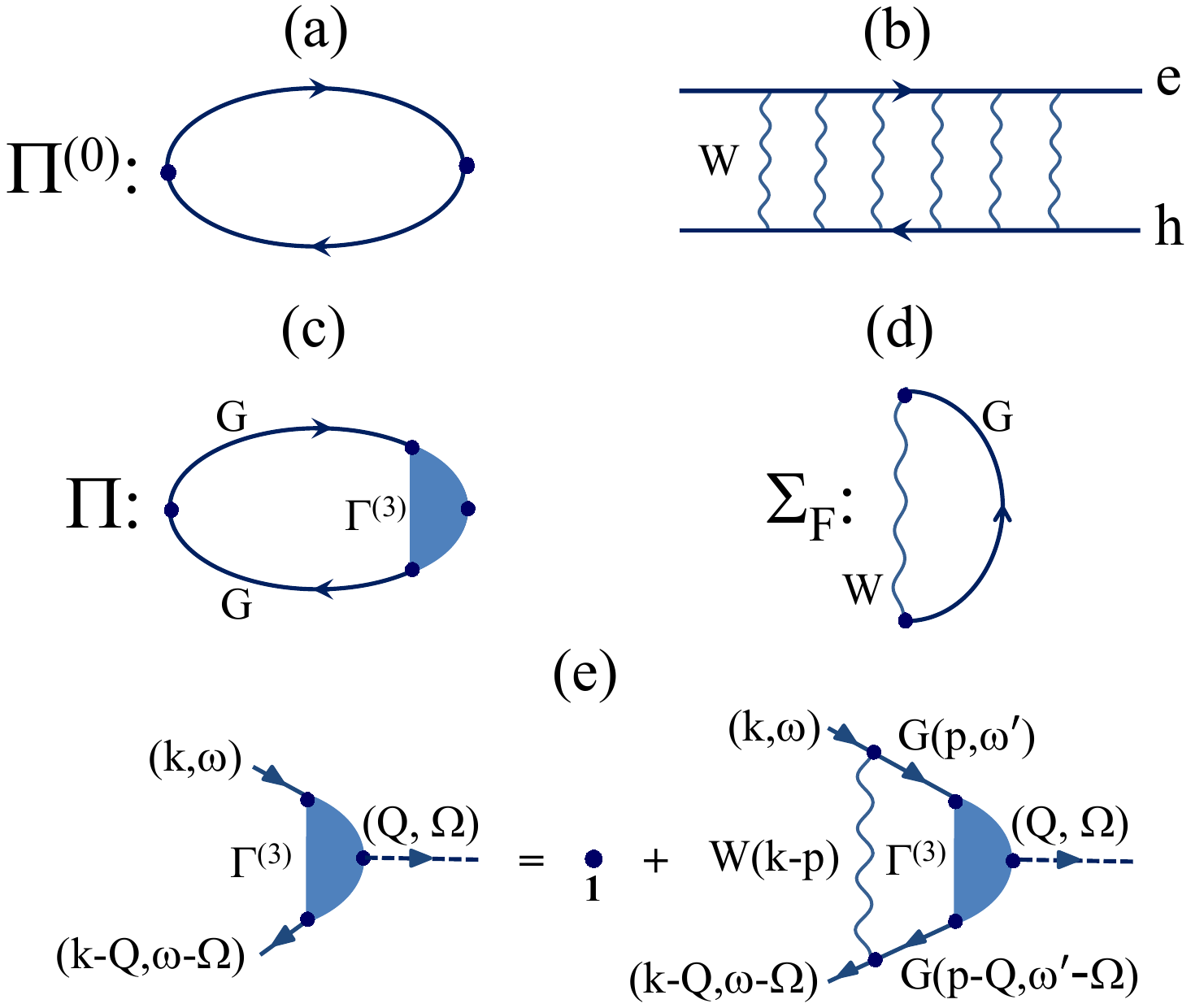}}
\caption{(a) The lowest order polarization diagram $\Pi^{(0)}$; (b) scattering of the electron-hole pair interacting via potential $W$; (c) expressing the polarization in terms of the dressed $3$-point vertex $\Gamma^{(3)}$;
(d) eelf-consistent Fock self-energy diagram; (e) Bethe-Salpeter equation for the vertex function $\Gamma^{(3)}(\textbf{k},\omega; \textbf{Q}, \Omega)$ dressed  by ladder diagrams.
Solid lines represent Green's functions in the HF basis and wavy lines represent static screened interaction,
$W(\textbf{k})$ (see \textit{Method}).}
\label{Fig1}
\end{figure}

The ideal gas zero-temperature expression for $\gamma_{\textrm{LD}}$ immediately follows from the
widely used random phase approximation (RPA) when the polarization is estimated from the lowest order
contribution, $\Pi^{(0)}$,  shown by the diagram (a) in Fig.~\ref{Fig1}.
However, using a simple analogy with the exciton problem in insulators, where Coulomb interaction
leads to radical changes in the behaviour of the particle-hole pair at the lowest energies and
formation of the bound state, one should expect that the value of $\gamma_{\textrm{LD}}$ in metals
is also strongly affected by the Coulomb force between the created electron and hole. The crucial difference,
of course, is that in metals interactions are screened at low-frequency.
The physics in question is described by the series of ladder diagrams shown in Fig.~\ref{Fig1}(b).
Within the field-theoretical framework, the problem boils down to calculating the polarization $\Pi$
using dressed 3-point vertex function $\Gamma^{(3)}$, see Fig.~\ref{Fig1}(c) and
Ref.~\cite{hedin1965}. In this re-formulation, summing up the series of ladder diagrams is equivalent
to solving the Bethe-Salpeter equation (BSE) \cite{BS1951} for $\Gamma^{(3)}$ presented schematically
in Fig.~\ref{Fig1}(e).

Summation of ladder diagrams for the polarization must be supplemented by simultaneous renormalization
of the electron Green's function, $G$, by considering non-crossing diagrams for the proper self-energy,
$\Sigma=\Sigma_F$, which is equivalent to the self-consistent Hartree-Fock (HF) approximation when
$G^{-1} = G_0^{-1} - \Sigma_F [G]$, see Fig.~\ref{Fig1}(d). Due to the charge neutrality of HEG
the Hartree contribution is absent. Thus, working in the HF basis not only guarantees that the
plasmon frequency obtained from the solution of  ${\rm Re} \, \epsilon (Q\to 0, \omega_p ) =0$ equation
is not shifted relative to the exact result $\omega_p = \sqrt{4 \pi n e^2/m}$, see Ref.~\cite{Pines},
but also radically simplifies calculations by eliminating the need for dealing with self-energy
type diagrams order-by-order.

In this work we apply the above field-theoretical setup (abbreviated as HF-BSE) to compute the system's polarization in a wide range of temperatures at metallic values of the Coulomb parameter $r_s$ and extract the LD coefficient from its imaginary part at small frequencies. This allows us to quantify the effects of pair scattering on top of the Fermi surface and reveal finite temperature effects as the system evolves from the correlated liquid state to the non-degenerate gas regime at $T \gg \varepsilon_F$. Finally, by extending simulations to arbitrary frequency and comparing to results obtained within a high-order diagrammatic scheme, we quantify the role of ``beyond HF-BSE'' diagrams and establish the typical functional form of the exchange correlation kernel at non-zero momentum and temperature within the HF-BSE approach, which is important for modeling of material dynamics within the time-dependent density functional theory (TDDFT), see, for instance, Ref.~\cite{Perdew2020} and references therein, and is considered to be one of the most important challenges in the modern theory of the electron liquid \cite{vignale}.


\section{Method}

Most finite-temperature many-body calculations for correlation functions proceed by first solving
the problem in the Matsubara representation \cite{Matsubara1955} and then applying the numeric analytic continuation
to the real-frequency axis in order to link theoretical results with experimental probes. However, the final
step is ill-conditioned and often distorts important spectral features (especially at low frequency)
even for very accurate imaginary-frequency data \cite{Goulko2017}. Instead, we rely on the diagrammatic Monte Carlo (diagMC) technique for computing $\Pi (Q,\Omega)$ from the ladder-diagram series directly on the real-frequency axis.
It was introduced in our previous work \cite{BSE2023} and here we apply it with minor technical modifications.

The key observation made in Refs.~\cite{LeBlanc2019,LeBlanc2020a,LeBlanc2020b} was that integration over Matsubara
frequencies and Wick rotation, $ i\Omega \to \Omega +i0$, of the final result to the real-frequency axis
for an arbitrary Feynman diagram formulated in terms of instantaneous interactions and ``coherent" Green's functions,
$ G^{-1} = i\omega_m -\epsilon (k)$, can be performed analytically; the corresponding automatic protocol
is called an algorithmic Matsubara integration (AMI). For analogous procedure in the real-time domain, see Refs.~\cite{FerreroRT,Ferrero2020}. The AMI protocol was recently used for computing the polarization
of the HEG \cite{LCHPT2022} by considering Taylor series expansion in the Yukawa potential \cite{Chen2019,Haule2022}.
However, the original formulation of the method was based on non-zero regularization parameter $\eta>0$ in the
Wick rotation $i\Omega \to \Omega +i\eta$ to avoid divergent statistical measures. On the one hand, simulations with non-zero $\eta$ introduce systematic bias. On the other hand, simulations with very small $\eta$, which are required to quantify the bias, are suffering from large statistical errors; this problem is especially severe for high-order diagrams and limits the accessible expansion orders.

An explicit procedure of taking the $\eta \to 0$ limit and eliminating singular statistical contributions from simple poles was proposed in Ref.~\cite{TTKP2021} and its first implementation for series of ladder diagrams was reported in Ref.~\cite{BSE2023}. While specifically designed and optimized for a limited set of diagrams, the $\eta=0$ scheme is very efficient and allows one to accurately compute contributions from expansion orders high enough for obtaining converged results or for using reliable resummation protocols for divergent series.

In all previous AMI simulations of the HEG  the diagrammatic series were formulated in terms of the Yukawa potential with some screening momentum $\kappa$ (expansion in terms of the bare Coulomb potential is ill-defined) and $\kappa$-counterterms. This is an exact procedure within the general shifted-action approach \cite{ShiftAct}, and one among infinitely many ways of formulating different diagrammatic expansions in terms of dressed, or renormalized, quantities. It allows one to expand in terms of an arbitrary potential $W(k)$ and counter terms $C(k)=1/V(k) - 1/W(k)$. If $W$ is chosen to be the Yukawa potential one faces the problem of selecting the ``best" $\kappa$ value. As famously advocated by P. Stevenson \cite{Stevenson}, the best choice is the one least sensitive to variations in $\kappa$, see also Ref.~\cite{Chen2019}. However, for the ladder-diagram series Ref.~\cite{BSE2023} established that the polarization is nearly independent of the choice of the screening momentum if $\kappa \sim \kappa_{TF}$, where $\kappa_{TF}$ is the Thomas-Fermi momentum. This observation suggests (and calculations explicitly verify the expectation) that equally accurate results are obtained for expansion in powers of $W(k) = V(k)/[1 - V(k)\Pi_{RPA}(k,\omega=0, T=0)]$, where $\Pi_{RPA}$ is the polarization of the ideal gas (the so-called Lindhard function). All our calculations were done using this form of $W(k)$.

Finally, we note that the self-consistent HF solution for the Green's function amounts to replacing the parabolic dispersion relation $\epsilon^{(0)}(k)=k^2/2m-\mu$ (counting energies from the chemical potential) with the HF expression,
$\epsilon(k)=k^2/2m + \Sigma_F(k)-\mu_F$, which preserves the simple pole structure of $G$. The chemical potential $\mu_F$ is self-consistently adjusted to keep the electron density fixed. Monte-Carlo simulations of the HF-BSE scheme described above are numerically exact and do not introduce any bias beyond the diagrammatic approximations.


\begin{figure}[t]
\centerline{\includegraphics[width=0.95\columnwidth]{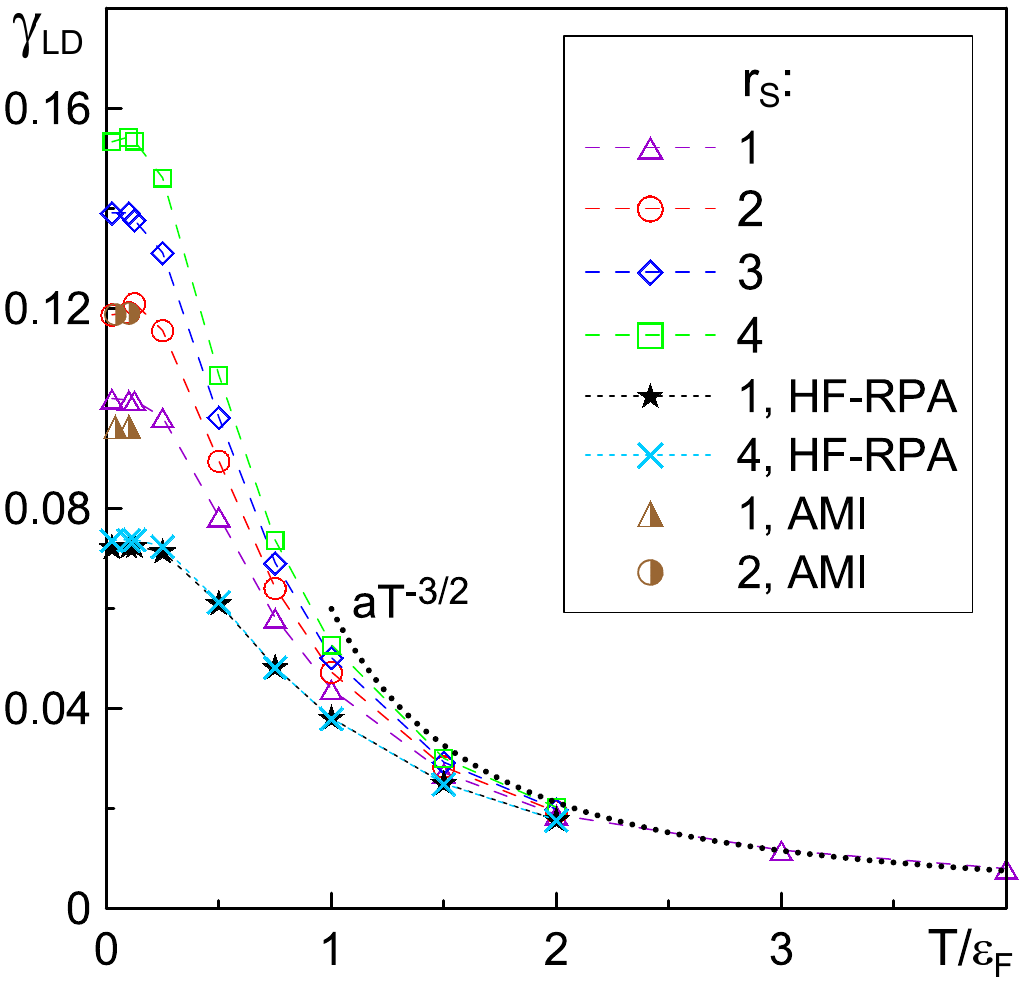}}
\caption{Landau damping coefficienst $\gamma_{\textrm{LD}}$ for different values of the Coulomb parameter $r_s$ as
functions of temperature within the HF-BSE approach. Simulations were performed at small value of the transfer momentum
$Q/k_F = 0.1$. For comparison, the two lowest curves are computed within the HF-RPA at non-zero temperature, i.e from the $\Pi^{(0)}$ diagram in Fig.~\ref{Fig1} (a), at $r_s=1$ and $r_s=4$, and the brown half-filled symbols show predictions of the high-order AMI simulations \cite{LCHPT2022} for $r_s=1$ (triangle) and $r_s=2$ (circle). Error-bars are within the symbol sizes. Dashed line is the asymptotic ideal gas law.}
\label{Fig2}
\end{figure}

\section{Landau damping}

Our results for LD coefficient are summarized in Fig.~\ref{Fig2}. The strongest renormalization of $ \gamma_{\textrm{LD}}$ takes place at low temperature, $T/\varepsilon_{\textrm{F}} \ll 1$, in the correlated Fermi liquid regime. The effect of Coulomb interactions is to increase $ \gamma_{\textrm{LD}}$ and it is getting more pronounced at larger values of $r_s$. One may wonder whether this behavior is the result of multiple re-scattering of the electron-hole pair on top of the Fermi surface or, at least partially, from working in the HF basis with the renormalized dispersion relation. To answer this question we also show results based on the $\Pi^{(0)}$ contribution, which is equivalent to the RPA in the HF basis (HF-RPA), see the lowest two curves in Fig.~\ref{Fig2}. Clearly, HF basis has little to do with the large
increase of Landau damping with interactions---close to a factor of two at $r_s=4$. Our results demonstrate that for the HEG system the RPA estimates for Landau damping are rather inadequate at low temperature and the problem is progressively more severe at larger $r_s$. Even at $r_s=1$ the $T=0$ value is significantly affected by the Coulomb scattering of the pair.

Non-zero temperature effects on $\gamma_{\textrm{LD}}$ become visible at $T/\varepsilon_{\textrm{F}} >0.1$, and around
$T/\varepsilon_{\textrm{F}} \sim 1$ we observe smooth crossover to the classical gas behavior. After significant drop from the $T=0$ values, all curves for different values of $r_s$ collapse at $T/\varepsilon_{\textrm{F}} > 2$ on the asymptotic ideal-gas high-temperature expression $\gamma_{\textrm{LD}} = a (\varepsilon_F/T)^{3/2}$ with $a=(2\sqrt{\pi}/3) \rho_F^{(0)}$.

\begin{figure}[t]
\centerline{\includegraphics[width=0.97\columnwidth]{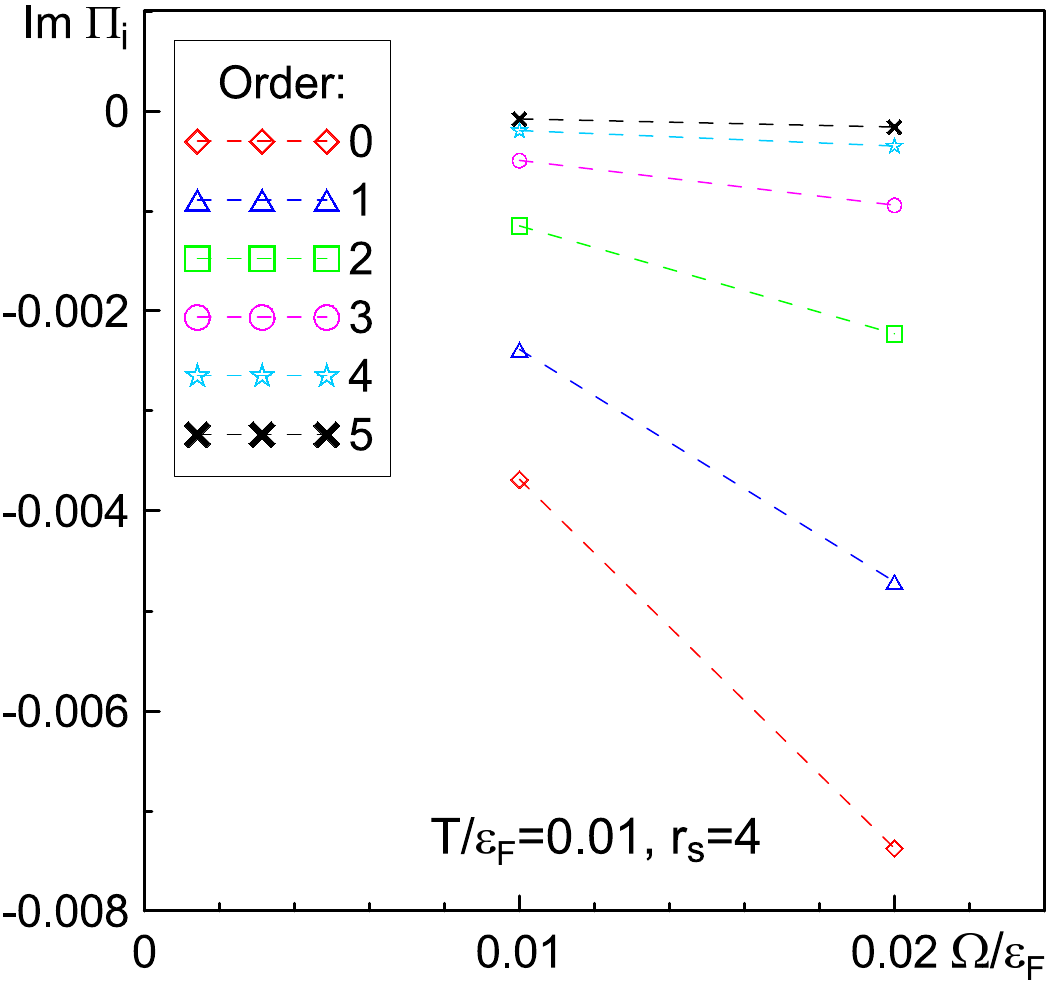}}
\caption{Partial contributions to $\textrm{Im} \Pi$ as functions of frequency at $T/\varepsilon_{\textrm{F}} = 0.01$, $r_s = 4$, and $Q/k_F = 0.1$. Each contribution represents a ladder diagram containing from $0$ (RPA-HF, see Fig.~\ref{Fig1}(a)) to five, see Figs.~\ref{Fig1} (c) and (e), interaction lines.}
\label{Fig3}
\end{figure}

To gauge the role of other diagram topologies on $\gamma_{\textrm{LD}}$, we also show in Fig.~\ref{Fig2} (brown half-filled symbols) predictions of the AMI simulations reported in Ref.~\cite{LCHPT2022} and conclude that for Landau damping contributions from these diagrams nearly cancel except at frequencies $\Omega > v_F Q$ were non-zero spectral density at $T=0$ is due to multiple $(e-h)$ pairs not included in the ladder-type HF-BSE scheme.

Finally, we would like to note that ladder-diagram series for $\gamma_{\textrm{LD}}$ are well converged. In Fig.~\ref{Fig3} we show partial contributions to $\textrm{Im} \Pi$ from diagrams of various orders---all curves (i) linearly extrapolate to the origin, see Eq.~(\ref{LDdef}) for the definition of LD, and (ii) clearly demonstrate that contributions beyond the fifth-order are negligible. The computational cost of the HF-BSE calculations is moderate; it takes about two days to produce all data in Fig.~\ref{Fig2} using a $16$-core desktop computer.


\section{Exchange-correlation kernel}

In the absence of precise data for frequency dependence of the HEG polarization, one of the most promising attempts to include effects of interactions on dynamics in comparison with the RPA solution is the exchange-correlation kernel approach. The kernel is formally defined by
\begin{equation}
K_{xc}(Q,\Omega, T) = \Pi^{-1}_{\textrm{RPA}}(Q,\Omega, T) - \Pi^{-1}(Q,\Omega, T) ,
\label{kernel}
\end{equation}
and this framework is used to suggest various functional dependencies of $K_{xc}$ constrained by exact sum rules and ground state energy dependence on density, see Ref.~\cite{Perdew2020} and references therein. In this work we examine the structure of $K_{xc}$ within the HF-BSE scheme directly from the accurate $\Pi$ and $\Pi_{\textrm{RPA}}$ calculations and demonstrate how it depends on frequency at finite value of momentum and temperature. This information cannot be deduced from sum rules and ground state energies and is considered to be very challenging \cite{vignale}.
\begin{figure}[H]
\centerline{\includegraphics[angle = 0,width=0.49\columnwidth]{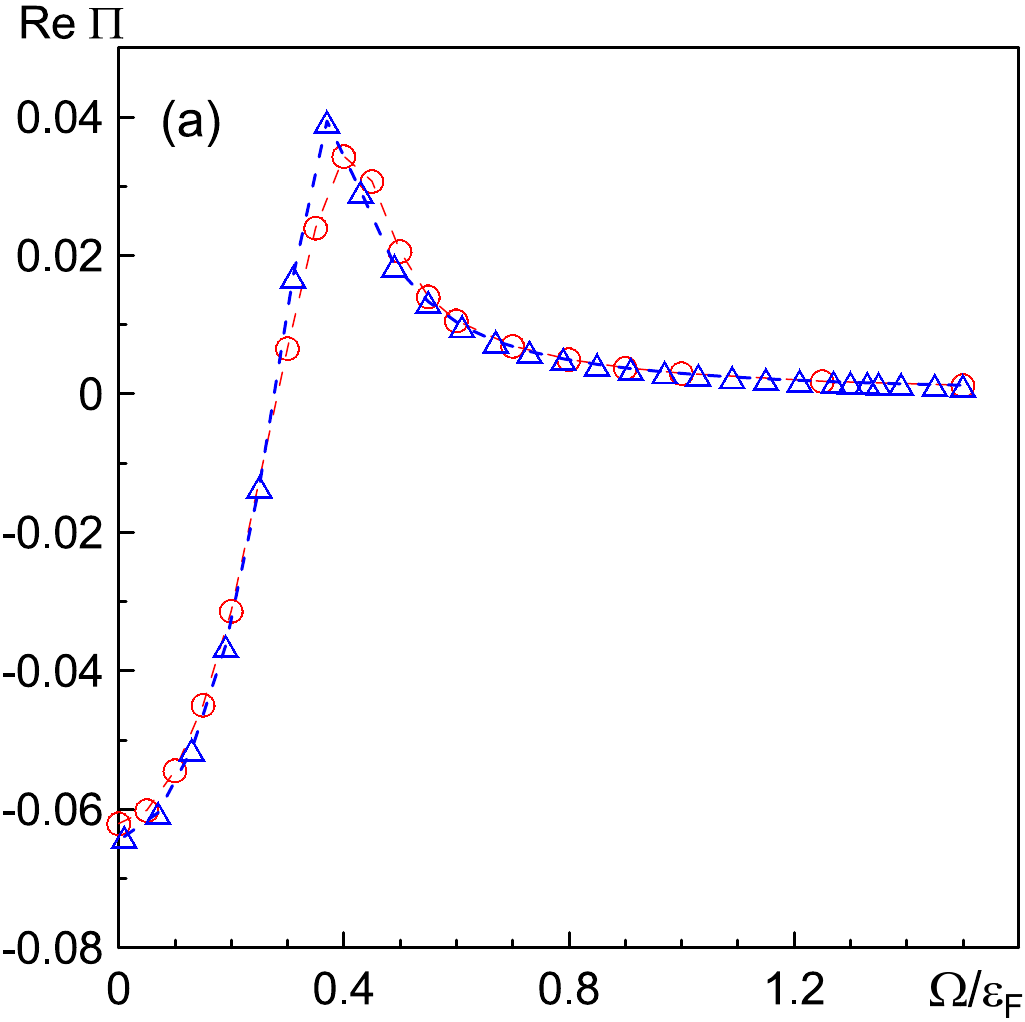}
            \includegraphics[angle = 0,width=0.49\columnwidth]{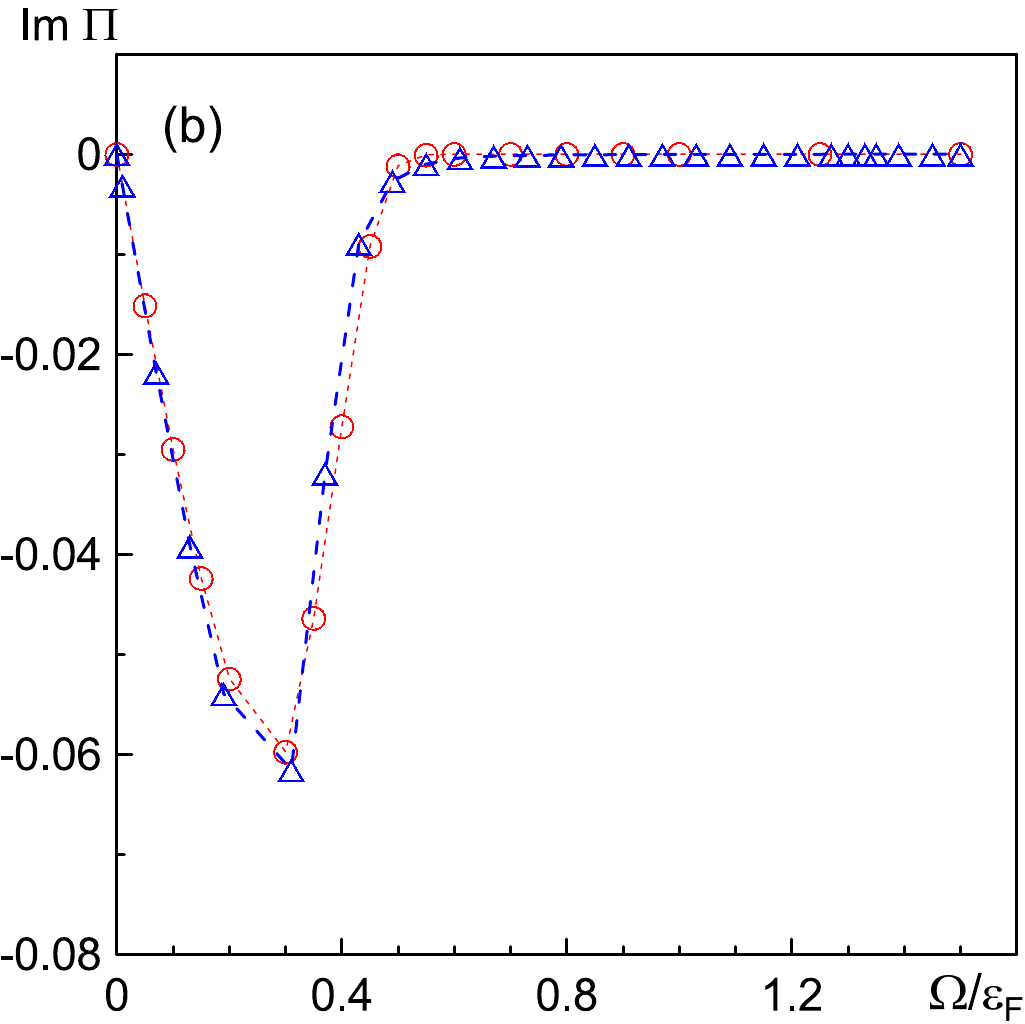}}
\centerline{\includegraphics[angle = 0,width=0.50\columnwidth]{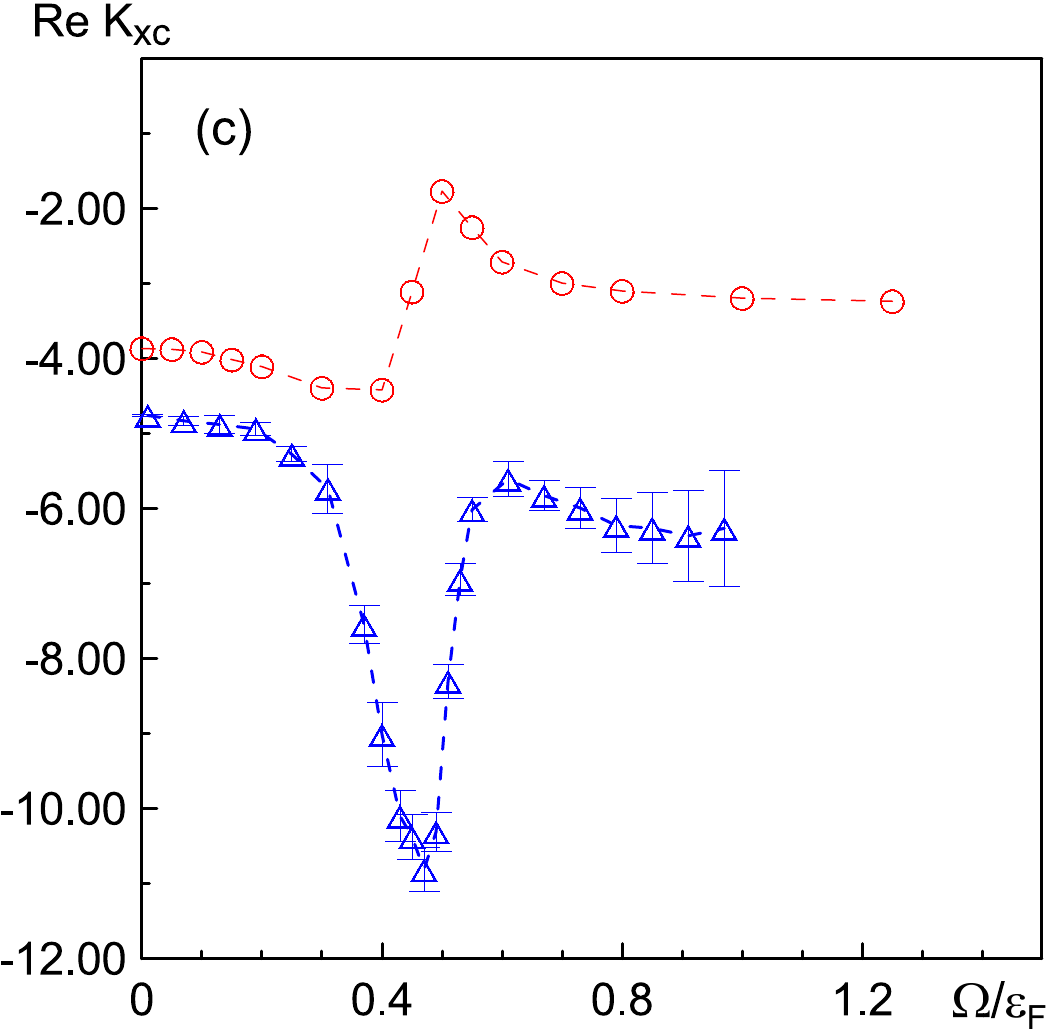}
            \includegraphics[angle = 0,width=0.49\columnwidth]{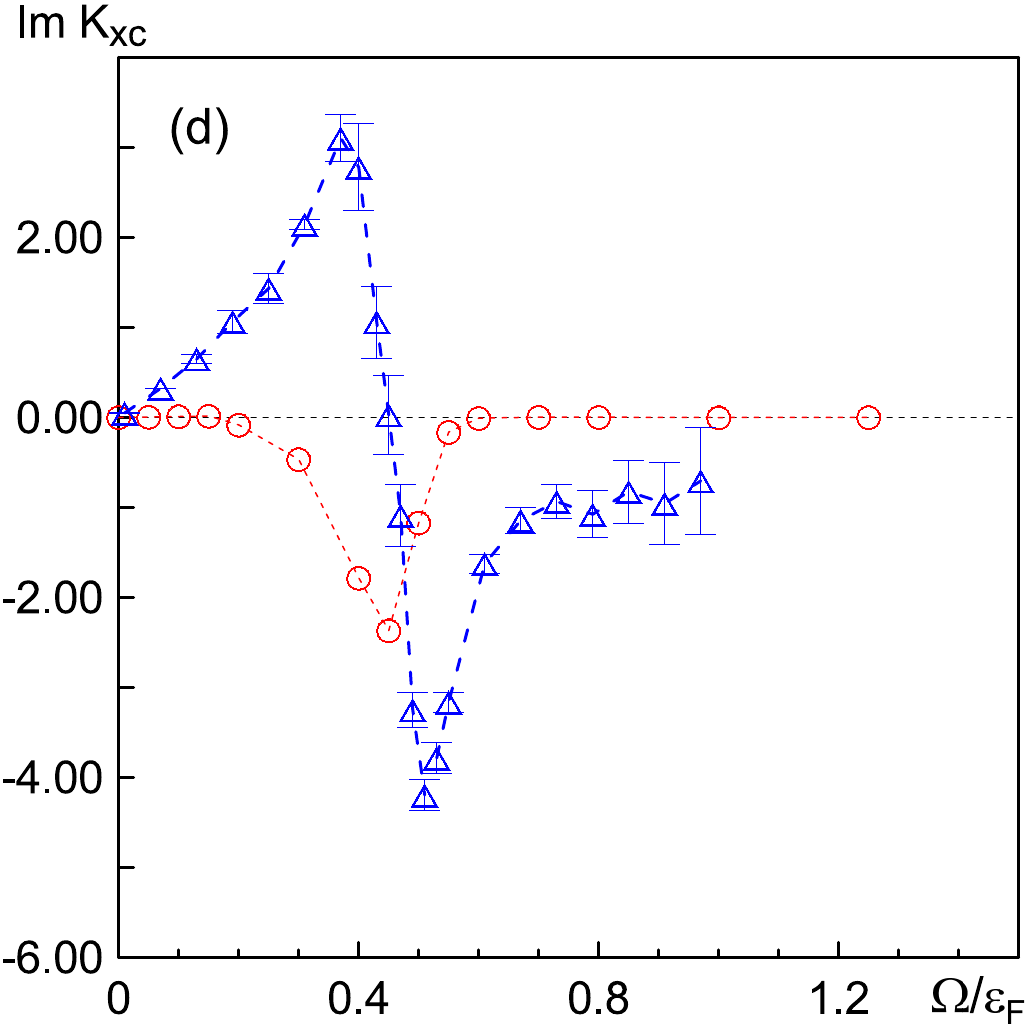}}
\caption{Real (left panels) and imaginary (right panels) parts of the polarization (upper panels)
and exchange-correlation kernel (lower panels) within the HF-BSE formalism (red circles) and high-order diagrammatic AMI scheme with finite regularization parameter $\eta$ (blue triangles) as functions of frequency at $T/\varepsilon_{\textrm{F}} = 0.1$, $r_s = 2$, and $Q/k_F = 0.2$. }
\label{Fig4}
\end{figure}

To obtain the kernel, we extend the frequency range of simulations all the way to $\Omega / \varepsilon_F \gg 1$ where $K_{xc}$ saturates to a plateau. In the upper panels of Fig.~\ref{Fig4} we present direct comparison between the
HF-BSE and high-order AMI results of Ref.~\cite{LCHPT2022} for polarization in the degenerate Fermi-liquid regime. It reveals that all AMI data are reproduced by the HF-BSE scheme very accurately, both for real and imaginary parts of $\Pi$. This further supports an earlier observation that at this level of comparison the contributions of high-order diagrams other than ladder type nearly cancel out and account only for small corrections.

However, since $K_{xc}$ is defined as the difference between the $\Pi^{-1}$ functions, even tiny variations between $\Pi$ and $ \Pi_{\textrm{RPA}}$ are magnified by orders of magnitude in regions where either function is small (note that $\textrm{Re} \Pi$ changes sign at $\Omega$ slightly below $v_F Q$). It is thus not surprising that HF-BSE and AMI results of Ref.~\cite{LCHPT2022} for $K_{xc}$ appear very different quantitatively, see lower panels in Fig.~\ref{Fig4}, while having the same characteristic shape in the vicinity of $\Omega / \varepsilon_F \sim 1$. The advantage of the HF-BSE scheme is that it is far more efficient (despite going to higher expansion orders for ladder-type diagrams to obtain converged results) and takes the $\eta \to 0$ limit exactly. As a result, it allows one to obtain precise data at high frequency where previous simulations failed because of exploding error bars.

The HF-BSE exchange-correlation kernel has the same key properties established in Ref.~\cite{LCHPT2022}: both real and imaginary parts feature a non-monotonic behavior with two extrema around $\Omega \sim v_F Q$ with the imaginary part violating causality at low frequency in contrast to early assumptions. On the one hand, these features reflect strong renormalization of the LD coefficient while preserving the total spectral weight of the $(e-h)$ continuum, which leads to several crossing points between the real and imaginary parts of $\Pi$ and $\Pi_{\textrm{RPA}}$, see Refs.~\cite{LCHPT2022,BSE2023}. On the other hand, they were missed in the phenomenological modeling of the kernel for lack of convincing evidence/arguments. Thus, we attribute the non-monotonic behavior of $K_{xc}$ to the processes in the $(e-h)$ continuum that can only be captured by high-order techniques; in particular, one has to account for Coulomb scattering of the $(e-h)$ pair non-perturbatively.

Clearly, accurate calculations of $K_{xc}$ are much more demanding than calculations of $\Pi$ or the charge response function $\chi = \Pi/\epsilon$, especially at frequencies where either $\Pi$, or $\Pi_{\textrm{RPA}}$, or both functions are getting very small. The latter situation takes place at high frequency when both functions go to zero while the difference between the two is approaching zero even faster. This is why observing the plateau in $K_{xc} =(\Pi - \Pi_{\textrm{RPA}}) / (\Pi \; \Pi_{\textrm{RPA}})$ is challenging numerically and requires a theoretical scheme where vertex corrections and self-energy renormalization of $G$ are properly balanced.


\section{Conclusions}

Using recently developed real-frequency Diagrammatic Monte Carlo technique \cite{BSE2023} for solving the Bethe-Salpeter equation in the Hartree-Fock basis at non-zero temperature for the ladder-type $3$-point vertex convoluted with two Green's functions, we studied the effect of Coulomb scattering on the electron-hole pair creation process  responsible for the Landau damping coefficient. We found that vertex corrections on top of the RPA result
(in the Hartree-Fock basis) in the degenerate ($T/\varepsilon_{\textrm{F}} \to 0$)
homogeneous electron gas strongly renormalize $\gamma_{\textrm{LD}}$ towards lager value
even at $r_s=1$, and this effect is getting more pronounced with increasing the Coulomb parameter.
On approach to the classical gas limit (at $T/\varepsilon_{\textrm{F}} > 2$) the vertex corrections become
negligible at all values of $r_s$ and $\gamma_{\textrm{LD}}$ follows the $\propto T^{-3/2}$ law.

Within the same theoretical framework we also studied the frequency dependence of the exchange-correlation kernel
and found that it features a highly non-monotonic behavior around $\Omega \sim v_F Q$ and lacks casuality
at $\Omega < v_F Q$, in agreement with previous results based on the full (but relatively low order)
diagrammatic expansion and non-zero parameter for regularization of poles.
These results demonstrate that starting from expansion in terms of the static screened interaction, solving
the self-consistent Hartree-Fock equations, and performing summation of ladder-diagram series for the polarization
we have an efficient computational scheme that captures most of physics right despite dealing with a limited
set of diagrams. Future work should explore the possibility of upgrading the scheme to the expansion
in terms of dynamically screened interaction when self-consistent Hartree-Fock is replaced with the self-consistent
$GW_0$ approximation.


\section{Acknowledgements}

This work was supported by the U.S. Department of Energy, Office of Science, Basic Energy Sciences, under Award DE-SC0023141.

\end{document}